\documentclass[%
 reprint,
 amsmath,amssymb,
 aps,
]{revtex4-2}

\usepackage{graphicx}% Include figure files
\usepackage{dcolumn}% Align table columns on decimal point
\usepackage{bm}% bold math
\usepackage{float}
\newcommand*{\affaddr}[1]{#1} % No op here. Customize it for different styles.
\newcommand*{\affmark}[1][*]{\textsuperscript{#1}}

\begin{document}

\preprint{APS/123-QED}

\title{A Heavy Ion Monitor on a Chip Based on a Non-Volatile Memory Architecture} % Force line breaks with \\

\author{
Dale Julson\affmark[1], Will Flanagan\affmark[2,3], Mike Youngs\affmark[4], \\Aidan Medcalf\affmark[1], Benedict Anderson\affmark[2], Sharanya Palit\affmark[2], Tim Hossain\affmark[1]\\
\affaddr{\affmark[1] Cerium Laboratories, Austin, TX, 78741, US}\\
\affaddr{\affmark[2] Department of Physics, University of Dallas, Irving, TX, 75062, USA}\\
\affaddr{\affmark[3] University of Texas, Austin, TX, 78712, USA}\\
\affaddr{\affmark[4] Texas A$\&$M Cyclotron Institute, College Station, Texas 77843,  USA}\\
}

\date{\today}

\begin{abstract}
The performance of a particle detector derived from nitride read-only memory (NROM) technology is evaluated, with immediate applications in space-based heavy ion radiation monitoring and detection. Irradiation exposures are performed using 40 MeV/u $^{78}$Kr and 10 MeV/u $^4$He particle beams at the Texas A\&M University Cyclotron Institute. The results show a strong sensitivity to high-Z heavy ions, and medium sensitivity to low-Z heavy ions.
\end{abstract}

\pacs{Valid PACS appear here}% PACS, the Physics and Astronomy
\maketitle

\section{Introduction}

A novel particle detector technology based on transistor structures traditionally used in non-volatile flash memory devices has recently been proposed~\cite{NAND_Heavy_Ion_Detector, flash_detector, NISoC, Tower_detector}. Such detectors offer significant size, weight, power and cost (SWaP-C) improvements over their traditional counterparts, including the ability to record particle interactions entirely unpowered. This is in part due to the standardized CMOS processes which these detectors are fabricated with, allowing for all requisite circuitry to be manufactured on a single ``system on a chip'' and for thousands of chips to be fabricated at once. In contrast to traditional silicon-based particle detectors in which electron-hole pairs are generated by ionizing radiation that are then measured as an induced current, in flash based detectors, these charges are separated via a ``write'' operation before any interactions occur. When ionizing radiation then moves through the sensitive element of the flash based detector, these charges are recombined, which can be measured at a later time as detailed in section II. These qualities have made this technology particularly applicable to space-based radiation monitoring, in which extremely low-power consumption circuits are advantageous due to the limited energy resources available. The need for such radiation monitoring technologies has been underscored by the recent exponential growth in the number of satellites in low earth orbit (LEO), as well as a renewed interest in human space exploration~\cite{satellite_population, space_exploration}. Both LEO satellites and crew are sensitive to the effects of ionizing radiation and would benefit greatly from the ability to accurately and quickly monitor the presence of such ionizing radiation~\cite{ion_beam_studies}. In fact, Radiation Monitoring Technology has routinely been deemed as one of the highest-priority technologies by NASA in regards to manned space exploration and sustained human activity beyond LEO~\cite{NASA_report}. This paper therefore proposes the Heavy Ion Monitor on a Chip (HIMoC), a flash memory based heavy ion monitor that addresses many of the shortcomings of previous technologies, such as improved reliability, while retaining high sensitivity and low-power requirements.

This paper is organized as follows. Section II provides a background overview of existing radiation monitoring technologies and an overview of flash memory devices. Section III discusses the experimental setup used in this paper. Section IV presents exposure data recorded at the Texas A\&M University (TAMU) Cyclotron in which two heavy ion species were tested. Section V presents a discussion of these results and a comparison of the advantages of NROM as a flash based detector technology. Section VI presents conclusions and next steps, while section VII provides acknowledgments.

\section{Background}

Traditional space-based radiation monitors and dosimeters have utilized a diverse array of technologies. Passive dosimeters, such as luminescence-based ones, can collect data without an active power source but lack the capability for real-time data monitoring~\cite{passive_dosimeters}. NASA’s active detectors, like the International Space Station Radiation Assessment Detector (ISS-RAD)~\cite{ISS_RAD} and the Hybrid Electronic Radiation Assessor (HERA)~\cite{time_pix}, provide real-time data monitoring but require a continuous power source to record data. In contrast, flash-based particle detectors are able to perform both passive particle detection and real-time monitoring. This is due to the unique architecture of the sensing element. 

\begin{figure}[ht]
% \centering
\includegraphics[width=0.4\textwidth]{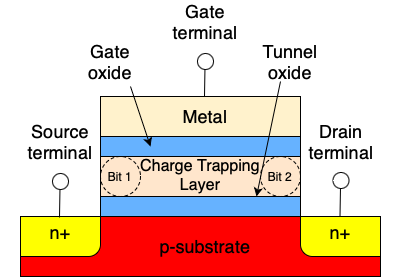}\\
\vspace{1em}
\includegraphics[width=0.4\textwidth]{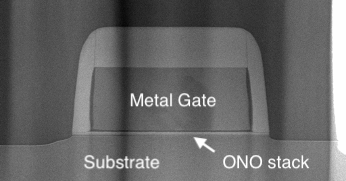}
\caption{(Top) A Representative cross section showing the gate stack up of a flash cell which consists of a charge trapping layer sandwiched between two oxide layers. ``Bit 1'' and ``Bit 2'' are only present in NROM devices, in which two separate localized pockets of charge are injected in each cell. (Bottom) A transmission electron microscopy (TEM) image of a SONOS device. Note the considerable difference in size between the metal gate and the ONO stack.\label{fig:NVM_Transistor}}
\end{figure}

\begin{figure}[]
 \begin{center} 
 \includegraphics[width=0.4\textwidth]{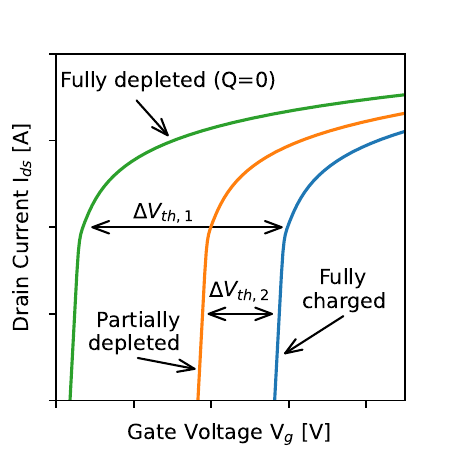}
 \end{center}
 \caption{The drain current versus gate voltage relationship of a flash cell. Shown in green is an uncharged cell which contains no trapped charge. When charge is injected into the charge trapping layer, this relationship is shifted upwards by an amount $\Delta V_{th,1}$, shown in blue. When ionizing radiation interacts with the cell, some amount of trapped charge is lost, causing a shift downward by the amount $\Delta V_{th,2}$, shown in orange. This relationship forms the sensing functionality of the cell.}
 \label{fig:Drain_current}
 \end{figure}

A typical flash memory element, shown in figure~\ref{fig:NVM_Transistor}, resembles a traditional MOSFET transistor with the addition of a charge trapping layer located beneath the gate of the device. When charge is injected into this layer, the threshold voltage of the device increases proportionally. This charge trapping material can consist of polysilicon metal, referred to as a ``floating gate'' (FG) device, or other dielectric material such as silicon nitride in which dangling bonds found in the bulk of the nitride layer act as localized charge traps~\cite{nitride_trapping}. Such nitride based configurations are referred to as ``SONOS'', named after the  Silicon-Oxide-Nitride-Oxide-Silicon gate stack. In FG devices, the charged metal forms an equipotential surface where trapped electrons are free to redistribute themselves, however in SONOS devices, the charges move very little after their initial injection into the charge trapping layer. The result of this is that charge can be trapped in a SONOS cell in different configurations depending on the charge injection method. If the charge is injected using Fowler-Nordheim (FN) tunneling, a homogeneous layer of charge is trapped along the entire length of the channel. If the charge is instead injected by channel hot electron injection, two distinct localized pockets located on either side of the channel can be formed. Such a configuration is referred to as Nitride Read-Only Memory (NROM) or ``Mirror-bit''~\cite{NROM1, NROM2}. The locations of the trapped charge within an NROM device are outlined by the dashed circles in figure~\ref{fig:NVM_Transistor} (labeled ``Bit 1'' and ``Bit 2''). The NROM transistor is the basic sensing element of HIMoC, and the interest of this paper. Each individual pocket of charge is able to be measured independently of the other by performing the measurement in ``reverse'', meaning the amount of charge above the source-channel interface (Bit 1) is measured by grounding the source, applying a drain voltage and gate voltage, and measuring the resultant drain current. Bit 2 can be measured similarly by swapping the drain and source voltages and performing the same routine. Figure~\ref{fig:Drain_current} shows the drain current - gate voltage relationship between a fully depleted cell with no charge in the charge trapping region (green), a fully charged cell (blue), and a partially discharged cell (orange). This relationship underlies HIMoC's sensing mechanism. All cells are initially programmed to be fully charged, causing the threshold voltage shift $\Delta V_{th,1}$. When ionizing radiation passes through the cell, some of the trapped charge is depleted from the charge trapping layer, causing threshold voltage shift $\Delta V_{th,2}$. By measuring the difference in threshold voltage between the fully charged cell and the partially depleted cell, a particle interaction can be detected. The detection mechanism requires no power to the device, and the interaction is recorded in perpetuity until the cell is reprogrammed.

 \section{Experimental Setup}

In order to test the sensitivity of HIMoC and validate its ability to effectively detect heavy ion particles, irradiation experiments were conducted with a commercial 1Gb NROM NOR flash device fabricated on a 90nm process node technology. The NROM cells were initially programmed using the internal write operation of the device (each localized bit in the charge trapping layer is fully charged). The threshold voltage for each bit was then measured and recorded by varying the wordline (gate) voltage of the cell from 0V until the bit ``flipped'' from ``0'' to ``1'' by using a custom readout assembly. Heavy ion exposures were then performed at the TAMU Cyclotron Institute using two different heavy ions beams, 40MeV/u $^{78}$Kr and 10MeV/u $^4$He. These two beams were selected to study the device response under high-Z heavy ions versus low-Z heavy ions. Each device was exposed only once, and new devices were used for subsequent exposures. After the devices were exposed, the new threshold voltage for each bit was recorded using the same process as before, and the resulting threshold voltage shifts were calculated.

 \begin{figure}[]
 \begin{center} 
 \includegraphics[width=0.4\textwidth]{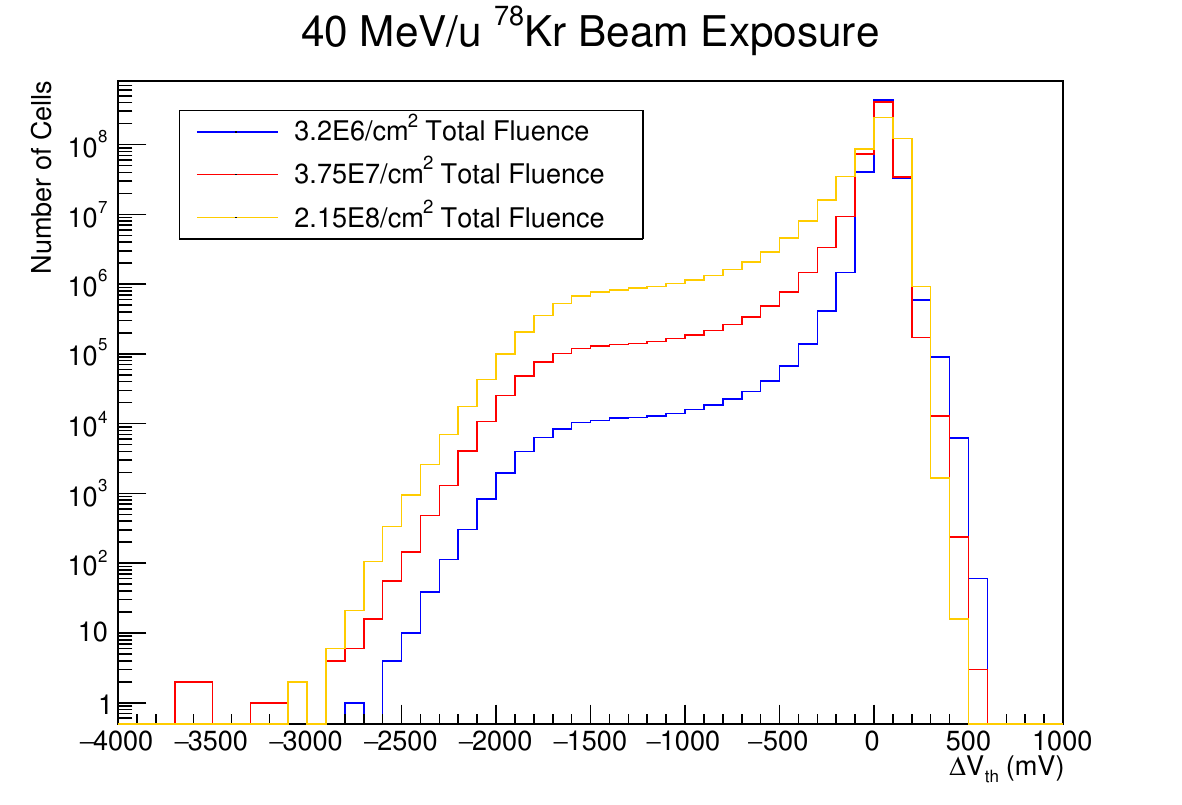}
 \end{center}
 \caption{Threshold voltage shift of each NROM device after 40MeV/u $^{78}$Kr beam exposure for varying fluence. Fluences recorded were 3.2E6 ions/cm$^2$ (blue), 3.75E7 ions/cm$^2$ (red), and 2.15E8 ions/cm$^2$ (yellow).}
 \label{fig:Kr78}
 \end{figure}

 \begin{figure}[]
 \begin{center} 
 \includegraphics[width=0.4\textwidth]{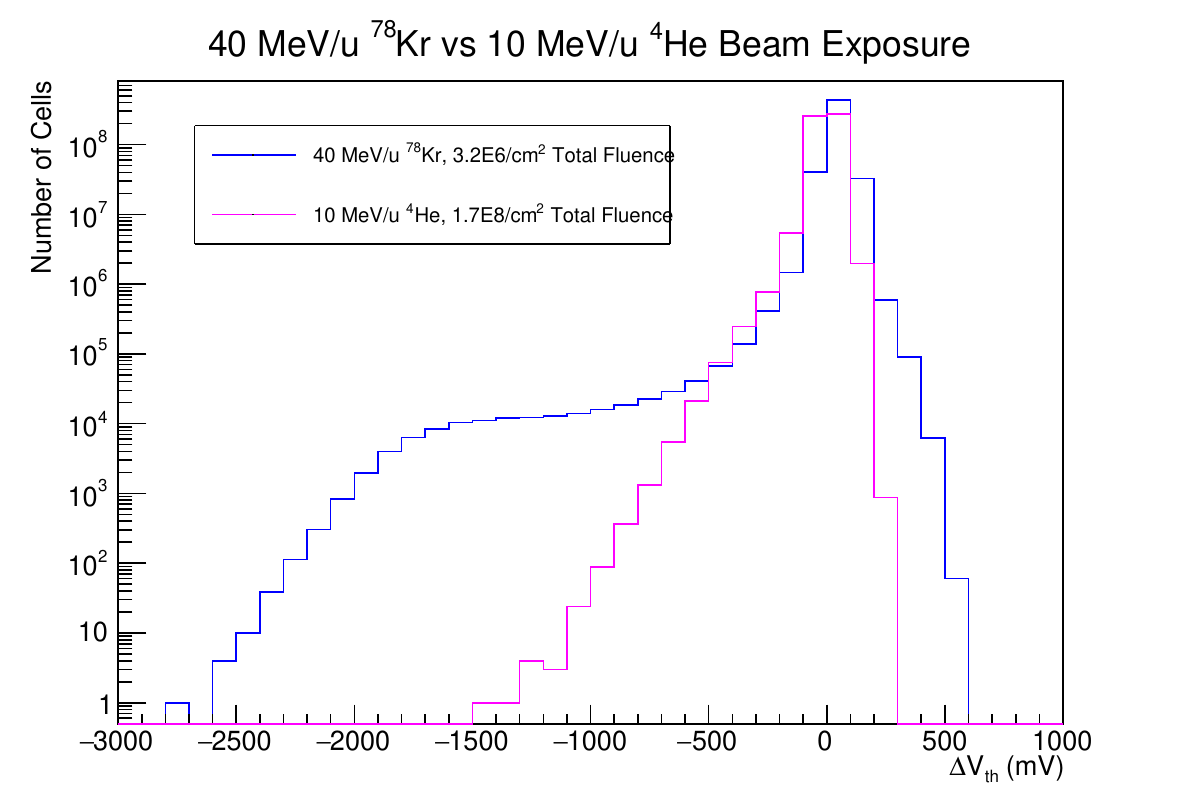}
 \end{center}
 \caption{Comparison of the NROM device response after 40MeV/u $^{78}$Kr beam exposure (blue) vs after 10MeV/u $^4$He (violet) beam exposure. The shape is noticeably different between the two isotopes, which can be attributed to the difference in linear energy transfer of the two ions.}
 \label{fig:He4}
 \end{figure}
 
\section{Results}\label{sec:Results}

Figure~\ref{fig:Kr78} shows the resulting shift in the threshold voltage for each bit (to be interpreted as $\Delta V_{\mathrm{th,2}}$ in figure~\ref{fig:Drain_current}) after $^{78}$Kr exposures of 3.2E6 ions/cm$^2$ (blue), 3.75E7 ions/cm$^2$ (red), and 2.15E8 ions/cm$^2$ (yellow) were performed. It can be observed that the resulting shape scales approximately proportional to the total fluence, but is otherwise unchanged. Figure~\ref{fig:He4} shows the results of the threshold voltage shift shape compared between the $^4$He exposure for a total fluence of 1.7E8 ions/cm$^2$ (magenta), with the $^{78}$Kr exposures of 3.2E6 ions/cm$^2$ (blue). The shape difference between these two distributions is significant, with $^{78}$Kr showing a much more pronounced threshold voltage shift despite absorbing an overall lower fluence. This is consistent with various results in the literature, for instance \cite{NAND_Heavy_Ion_Detector}, in which the shape of the threshold voltage shift distribution is dependent on both the particle species and linear energy transfer (LET).
 
\section{Discussion}

It has been previously reported that SONOS based devices, when programmed using FN tunneling so that there is a uniform charge distribution across the entirety of the charge trapping region, are less sensitive to heavy ion radiation than their FG counterparts~\cite{sensitivity}. This is due to the fact that in a FG device, the charge present in the charge trapping layer is able to freely move and redistribute itself. When a heavy ion strikes a flash cell, it is theorized that the discharge mechanism is a temporary ``transient conductive path'' formed in the channel oxide layer which discharges the charge trapping layer~\cite{conductive_path}. For a FG device, this discharge is able to occur uniformly across the entire layer, regardless of where the cell was struck. For a SONOS device however, the discharge only occurs in the local region in which the heavy ion struck the device. In order for the effective threshold voltage of the SONOS cell to be lowered, it can therefore require multiple hits occurring in specific locations across the channel that connect the source and drain~\cite{modeling_SONOS}. This improved sensitivity of FG cells in comparison to SONOS configurations is not without its drawbacks however. It has been shown that heavy ion strikes in FG cells can create small, permanent leakage currents which percolate through the channel oxide, causing the cells to discharge over time, even when not exposed to radiation~\cite{Rad_effects_NROM}. Similar effects are not observed in SONOS devices~\cite{data_retention}. This limits the usefulness of FG cells in applications in which a detector should be used more than once. An NROM device comparatively combines the advantages of both of these architectures. The charge in the NROM charge trapping layer is already highly localized in two discrete packets, meaning the device does not require multiple hits to lower the effective threshold voltage, but instead only one hit along this packet in order to affect the threshold voltage. The result is a similar sensitivity to FG devices but with better durability, due to the non-conductive nature of the charge trapping layer. 

\section{Conclusion}

A novel heavy ion radiation monitor on a chip has been proposed and validated. The immediate proposed application for this technology is in space-based radiation monitoring for both satellites and human astronauts. It is proposed that by stacking together multiple HIMoC chips, either through custom PCBs or in advanced heterogeneous 3D packaging, particle tracking can be performed. Additionally, future efforts propose reducing the the external packaging of the chip to maximize sensitivity to heavy ion radiation.

\section{Acknowledgements}

This work was supported in part by a Phase II contract from the U.S. Air Force Small Business Technology Transfer program (contract number FA864922P0832), the Texas A\&M University Cyclotron Institute  (DOE DE-FG02-93ER40773, NSF PHY-2051072), and the University of Dallas Donald A. Cowan Physics Institute. Data from contract number FA864922P0832 approved for public release; distribution is unlimited. Public Affairs release approval \#AFRL20240363.

%\newpage

\end{document}